\documentclass{article}
\usepackage[margin=2cm]{geometry}
\pdfoutput=1 

\usepackage{microtype}

\providecommand{\tightlist}{%
  \setlength{\itemsep}{0pt}\setlength{\parskip}{0pt}}

\usepackage{amsmath}
\usepackage{amsfonts}
\usepackage{amsthm}
\usepackage{booktabs}
\usepackage{xspace}

\usepackage{graphicx}
\usepackage[tiny,compact]{titlesec}
\usepackage{url}
\usepackage{paralist}
\usepackage{enumitem}
\usepackage{natbib}

\usepackage{hyperref}

\newcommand{\nconf}{63\xspace}
\newcommand{\narea}{25\xspace}
\newcommand{\ntopconf}{17\xspace}

\newcommand{\ntotaldblp}{82427\xspace} \newcommand{\nmatched}{7313\xspace}

\title{Popularity of arXiv.org within Computer Science}
\author{Charles Sutton and Linan Gong \\
{\normalsize School of Informatics, University of Edinburgh, UK} \\
{\normalsize \url{csutton@inf.ed.ac.uk}}}

\begin{document}
\maketitle

Research 
depends on rapid dissemination of results,
and computer science researchers have been
naturally attracted to the idea of placing their
papers on the Web.
Indeed, in 1998 the Computing Research Repository (CoRR)
was established as a partnership between
the ACM, arXiv.org, and others as an online repository for
research papers in computer science
\citep{halpern2000corr}.
However, CoRR seems to have had fairly limited
popularity for the first decade of its existence
(Section~\ref{results}), with researchers preferring
to post e-prints on their individual
web sites \citep{mccallum:cora,giles98citeseer}.
At the same time, we are seeing a budding revolution
in the way that publication and peer review is conducted in computer
science. Some researchers have argued for all submissions and reviews
to be public \citep{lecun:manifesto}, and for four years
the International Conference on Learning Representations (ICLR) has followed this
model successfully, increasing from 67 submissions in 2013 up to 647
submissions in 2017.	\footnote{The text and reviews of all accepted and rejected ICLR 2017 submissions
are available at \url{https://openreview.net/group?id=ICLR.cc/2017/conference}.}
To support experimentation with new and open
reviewing models, the OpenReview system \citep{openreview:web} was designed specifically to allowing conferences to configure a wide
range of reviewing and dissemination policies
and levels of openness; a good overview of the debate about reviewing models is provided
by the accompanying paper \citep{openreview}.

An important aspect of how research is disseminated
in computer science
is \emph{whether}, \emph{where}, and \emph{when} papers are made
electronically available.
For many years, researchers have made papers
available via individual home pages,
and some researchers use or are mandated
to use institutional repositories.  
However, public e-print services like CoRR/arXiv
have the advantage of allowing
researchers
to receive automatic notifications when new relevant papers are uploaded (and incidentally,
also of facilitating the study of the literature
in aggregate, as in the current paper).

There is also the question of when papers should
be made public, i.e. immediately after the authors
judge them
ready for public consumption, which
we will call \emph{prepublication}, or only after
they have been accepted by a peer reviewed
conference or journal.
Advocates argue 
that
 prepublication can have a dramatic effect on
  the rate of change in a field. For example, in deep learning, it is
  now impossible to stay up to date without following submissions in
  arXiv closely. 
In some well-known cases, by the time an influential
paper has been presented in a conference,
follow-on work has already appeared,
and indeed, is sometimes presented
in the same conference session.
However, an important disadvantage of 
pervasive prepublication is that it significantly
complicates double-blind reviewing, 
as a centralized e-print repository can send
alerts to many
potential reviewers.
Many research communities within computer science have had multi-year discussions
about the tradeoffs among e-print servers, prepublication, and reviewing models,
with concomitant town hall meetings at conferences,
blog posts, and so on.\footnote{Examples include
\url{https://gridworld.wordpress.com/2015/07/20/arxiv-heaven-or-hell/}, 
\url{http://hunch.net/?p=1086}, and \url{https://chairs-blog.acl2017.org/2017/03/02/arxiv-and-double-blind-reviewing-revisited/}.}
As the comments on some of these blogs indicate,
the debate among researchers can be spirited,
even if it is carried out with
only the most limited of evidence.
Although all indications are that computer scientists
are prepublishing their articles more and more,
there is little information about the size of this trend
or how it varies across areas of computer science.

We aim to quantify the extent to which computer scientists deposit papers in 
centralized e-print repositories, and whether these e-prints are preprints
that are made public before or during the review process, or whether e-prints
are typically posted only upon the paper's acceptance into a peer-reviewed venue.
Specifically, we measure the percentage of 
computer science publications that have been
posted as e-prints to arXiv.org.
We consider all papers in the past
decade that have been
published in \nconf of the most selective conferences
across computer science, and matched their
metadata to the arXiv preprint server.  
Our main findings are:

\begin{itemize}
\tightlist
\item
  Usage of arXiv.org has risen dramatically among the most selective conferences in computer science.
  In 2017, fully \textbf{23\% of papers} 
  had e-prints on arXiv.org, compared to only 1\% of papers ten years
  ago.
\item
  Areas of computer science vary widely in e-print prevalence. 
 In theoretical
  computer science and machine learning,
 over 60\% of published papers have arXiv e-prints.
 In other areas, the arXiv usage is
 essentially zero.
  In most areas, arXiv usage is rising.
\item
 Many researchers use arXiv for posting preprints.
 Of the 2017 published papers with arXiv e-prints,
 \textbf{56\% were preprints} that were posted
either before submission or while the
review process was ongoing.
\end{itemize}

Overall, these results suggest that we are reaching
a tipping point in the popularity of centralized
e-print repositories.
This has implications for both researchers and computing practitioners.
For practitioners, following
arXiv feeds has now become an important tool for tracking recent
literature in many areas of computer science. 
For researchers, these results
highlight the urgency of having an explicit discussion about how prepublication should affect community norms for
dissemination, reviewing, and publication (Section~\ref{policy}). 
We may well have already reached a point
where preprints are here to stay.

\emph{Terminology: e-prints, preprints, and all that.} 
We use the term \emph{e-print}
to indicate a version of a research article
that is publicly placed online by the authors
directly.
This includes e-prints in centralized repositories
like the arXiv,  in repositories of individual
institutions, or on researchers' individual or
group Web pages.
A \emph{centralized e-print repository} 
is an internet service that stores
and distributes the e-prints of many
different research institutions and
journals, such as arXiv.org, bioRxiv, HAL, and so on.
A \emph{preprint}, for us, is an e-print that is made public
before the paper has been accepted into a peer-reviewed venue; of course, some preprints
never appear in a peer-reviewed venue.
Although the practice of e-print repositories
has precursors in processes for exchanging
paper manuscripts 
\citep{kling05}, in this analysis we are only concerned
with preprints that appear electronically.
So all preprints are e-prints, but not vice versa.
By \emph{prepublication} we mean the practice of using
preprints to disseminate research.

\section{Context}

Researchers in computer science are in the midst
of a lively, multi-year debate about reviewing
and publication models. Unlike other sciences,
in computer science, the primary focus of publishing
has been in refereed conferences, which typically
have acceptance rates of between 10\%-30\%.
This has been perceived to be a faster publication
mechanism than journals, but more recently researchers
have expressed frustration with the delay
in dissemination that can result when papers are
rejected multiple times from conferences, waiting
several months each time for the next conference deadline.
This ``conference ping-pong'' can affect
even groundbreaking papers
\citep{lecun:manifesto}, indeed, perhaps such
papers are especially susceptible,
as they are by nature
more difficult for peer reviewers to understand
and appreciate.
Concerns about reviewing and publication models
have led to discussions across areas of computer
science,
workshop such as one on publication models
 at the NIPS conference in 2009
 and
one on peer reviewing and publication
models at ICML 2013\footnote{\url{https://sites.google.com/site/workshoponpeerreviewing/}}.
It has also led to the foundation of hybrid venues
which have journal-style rolling deadlines but include
conference publication, such as
the Proceedings of the VLDB Endowment,
the Transactions of the Association of Computational
Linguistics, and the journal track of the
European Conference on Machine Learning and Principles and Practice of Knowledge Discovery in Databases.

Posting preprints to centralized e-print
repositories provides a way to decouple
dissemination from peer review, allowing
papers at risk of conference ping-pong 
the opportunity to be read and noticed more quickly.
It also provides advantages to researchers in smaller
groups and less developed countries,
allowing them fast access to the most cutting-edge
literature and also the ability to disseminate
their own research results quickly.
Indeed,
Ginsparg notes that twenty years after
the creation of arXiv.org: ``I still
receive messages reporting that the system
provides to them more assistance than any
international organization''
\citep{ginsparg11twenty}.

Leading conferences in computer science
have contradictory
policies regarding the use of centralized e-print
repositories and preprints. 
In theoretical computer science,
STOC and FOCS have explicitly encouraged authors
to make full versions of their paper 
available as e-prints.\footnote{Sources: \url{http://acm-stoc.org/stoc2017/callforpapers.html} and \url{https://focs17.simons.berkeley.edu/cfp.htm}}
Notably, these venues do not use double-blind review.
In machine learning, NIPS and ICML have
allowed authors to place preprints online
for several years.
On the other side, AAAI 2017 and SIGIR 2017
discouraged authors from placing preprints online during the review process,\footnote{Sources: \url{http://www.aaai.org/Conferences/AAAI/2017/aaai17call.php} and \url{http://sigir.org/sigir2017/submit/call-for-full-papers/}}
but for AAAI, many authors appear to have been unaware 
of this request (Table~\ref{tbl:top_venues}).
ACL 2017 required that authors declare
at the time of submission whether authors
had placed a preprint on arXiv;\footnote{\url{http://acl2017.org/calls/papers/\#multiple-submission-policy}} papers with
undisclosed arXiv preprints 
were
rejected without review.
An interesting compromise has been used by PLDI\footnote{\url{https://pldi18.sigplan.org/track/pldi-2018-papers\#FAQ-on-Double-Blind-Reviewing}}, which allows submitted
papers to have arXiv e-prints as long as they were posted to arXiv well before
the conference submission deadline, in order to avoid an arXiv announcement with a large
number of papers just after the conference deadline attracting
the notice of potential reviewers.
Similarly, PLDI discourages publicizing e-prints to social media
and mailing lists during the review process.

Therefore, a study on the usage of centralized
e-prints and preprints in computer science
has several policy objectives.
First, this study aims to provide an opportunity for researchers
and research communities
to reflect on their own policies and norms
for centralized e-print repositories and preprints in light of
the experience of other areas.
 For example, areas with less of 
a tradition of the use of centralized e-print
repositories might be inspired (or perhaps repelled!) by areas that do.
Second, this study aims to provide evidence to inform the 
current discussion of double-blind reviewing
systems to better accommodate prepublication models.
Measures of prepublication can help to inform
the debate
on whether to disallow 
preprints at double-blind conferences,
whether to discontinue double-blind reviewing
to better accommodate preprints, or whether and how
to design a hybrid system that incorporates both.
Finally, this study aims to provide insight into future prospects
of open reviewing models across computer science.
Open review requires preprints by definition,
so if computer
  science researchers are increasingly coming to accept prepublication,
this bodes well for the possibility of 
current experiments in open review to
  gain traction in the community.

Outside of computer science, \cite{lariviere14} have previously studied
arXiv usage across all sciences, ranging from mathematics and physics to chemistry and biology.
They follow a similar methodology to ours, matching journal publications from Web of Science to e-prints
on the arXiv. While they do consider computer science, they only consider papers 
in computer science journals rather than refereed conferences, and hence does not consider
perhaps the most important route for research dissemination in computer science. 
Finally, this study stops in 2011; as we will see, computer scientists' usage of arXiv has changed dramatically in the last six years.
They find that the percentage of journal papers on arXiv
varies substantially across disciplines, which
is consistent with the intra-disciplinary results
for computer science that we report here.

\section{Methods}\label{methods}

The simplest method for assessing the increase
in e-print usage is to examine the publicly
available statistics from arXiv.org\footnote{\url{https://arXiv.org/help/stats/2016_by_area/index}}, which
indicate that the number of arXiv submissions is increasing across subject areas of computer
science. But this analysis alone cannot indicate
whether the number of arXiv submissions is
increasing because the percentage of researchers
posting e-prints is growing, or simply due
to the overall growth in the number of computer science researchers.
Furthermore, raw submission counts cannot be used to
  meaningfully compare areas, 
 because different areas of
  computer science have different sizes.
Finally, submission counts cannot on their
own tell us whether researchers are using arXiv
for prepublication or whether they 
post to arXiv after the review process is complete.
Therefore, our method is instead to 
collect the list of all
published papers from top conferences across 
computer science, and match the published
paper to a corresponding arXiv submission.
This allows us to measure what percentage
of papers are submitted as arXiv e-prints.

\begin{table}
	\centering
	\begin{tabular}{ll}
\toprule
                       Area &                     Conferences \\
\midrule
  Algorithms and complexity &                STOC, SODA, FOCS \\
    Artificial intelligence &                     AAAI, IJCAI \\
     Computer architectures &             MICRO, ISCA, ASPLOS \\
               Cryptography &               EUROCRYPT, CRYPTO \\
                Data mining &                             KDD \\
                  Databases &                   PVLDB, SIGMOD \\
          Design automation &                      DAC, ICCAD \\
          Embedding systems &              RTSS, RTAS, EMSOFT \\
                   Graphics &                SIGGRAPH (+Asia) \\
                        HCI &              UbiComp, CHI, UIST \\
 High performance computing &                   SC, ICS, HPDC \\
      Information retrieval &                      WWW, SIGIR \\
     Logic and verification &                       CAV, LICS \\
           Machine learning &                      NIPS, ICML \\
                Measurement &                 SIGMETRICS, IMC \\
           Mobile computing &        MobiSys, MobiCom, SenSys \\
                        NLP &           ACL, HLT-NAACL, EMNLP \\
                 Networking &          NSDI, INFOCOM, SIGCOMM \\
          Operating systems &      USENIX, SOSP/OSDI, EuroSys \\
      Programming languages &                      POPL, PLDI \\
                   Robotics &                 RSS, ICRA, IROS \\
                   Security &  CCS, IEEE S\&P, USENIX Security \\
       Software engineering &                  ASE, ICSE, FSE \\
                     Vision &                ECCV, CVPR, ICCV \\
              Visualization &                   IEEE VIS + VR \\
\bottomrule
\end{tabular}

	\caption{Conferences included in this study, listed by research area.}\label{tbl:conferences}
\end{table}

We focus on papers published in the most
selective conferences in computer science,
both as a way of narrowing the scope of the study,
and also as a way to assess practices among
 leaders in the field. A list of such
conferences has been compiled by 
CSRankings \citep{csrankings:web}, a recent effort
to create a metrics-based ranking of worldwide
computer science research departments.
CSRankings has collected a list of research
areas in computer science, roughly corresponding to
ACM SIGs, but filtered to ensure that each area
has a sufficiently broad research community.
More details about the selection of the 
CSRankings list is available at 
\url{http://csrankings.org/faq.html}.
We retrieved the list in from the CSRankings
GitHub repository in December 2016.
We made two changes to the list from CSRankings:
first, we excluded the bioinformatics area,
as much prepublication in that area seems 
to occur on bioRxiv (\url{http://biorxiv.org/}),
which is not included in this study,
rather than arXiv.org.
Second, we split the KDD conference into a separate
"data mining" area apart from ICML and NIPS, as 
the two areas had different prepublication behaviour.
This resulted in a list of \nconf conferences,
spanning \narea areas of computer science,
shown in Table~\ref{tbl:conferences}.

Proceedings of the 
conferences ISMB, SIGGRAPH, SIGGRAPH Asia,
IEEE VIS, and VR are published as special issues
of journals, so we include in our study the papers
from the relevant issues of the associated journals.
Short papers in ASE and ICSE are omitted.
Code to handle this processing was 
provided by CSRankings.
We collected the list of published papers from
each of the conferences from 2007-2017 using DBLP (\url{http://dblp.org/}).
The venue names from DBLP were normalized using
rules provided by CSRankings.org.
To collect the list of arXiv e-prints,
we retrieved the metadata of all papers
from the \texttt{cs} and \texttt{stat} subjects,
a total of 177\,129 e-prints.
The \texttt{stat} subject was included because
many machine learning papers on arXiv
are	posted to the \texttt{stat.ML} category.
For arXiv, the OAI2 interface does not provide individual authors
separately, so we heuristically separated author names by commas and
the word ``and''.

For each published paper in the conferences
considered, we attempt to determine
whether there is a submission on arXiv
that matches the paper metadata from DBLP.
Citation matching has been an active research area
for many years
\citep{lawrence1999autonomous,mccallum2000efficient}, but we use a simple approach, 
both because our metadata is likely to be  
more accurate than citation data, and also 
because 
we want a conservative measure. Therefore we 
use the following record linkage heuristic.
We consider
that a citation in DBLP matches an e-print on the arXiv
when the titles are exactly identical and at least
one token is shared between the author fields
of the two records. 
Overall there are \ntotaldblp papers in total reported on DBLP
from the \nconf conferences for the years considered
in this study. Of these, according to our matching heuristic,
\nmatched of these papers have arXiv e-prints.

We estimate the effectiveness of our record linkage heuristic
via the standard measures of pairwise precision and recall.
First, for precision,
we randomly sampled 25 DBLP papers for
which our record linkage heuristic did find
a match.
Of the 25 matched papers, we compared the 
arXiv e-prints and DBLP papers manually to check
if they presented substantially the same major results, subject to minor expansions such as
inclusion of proofs and so on.
We found that all of the 25 matches from the heuristic
were indeed correct, so that the precision
of the heuristic is as hoped nearly 100\%.
Second, to estimate recall, 
for each of the papers in DBLP that were not matched by our heuristic,
we found the closest arXiv e-print measured by the
average of the Jaccard similarities of the title
and author fields across the two records.
In 74644 cases, the closest arXiv e-print had
Jaccard similarity of
less than 0.5; we assume that all such arXiv
e-prints do not match the corresponding DBLP record. 
For the remaining 470 papers in DBLP, we randomly sampled 25 for manual
comparison, and found that 23 of
the 25 closest arXiv e-prints were also true matches.
This yields an estimate of the recall of our 
heuristic as 94\%, which is high enough to not have a serious
impact on our subsequent analysis. Although it would be
possible to relax the heuristic a bit to increase
the recall while suffering only a slight drop
in precision, we choose not to do this
for the sake of simplicity.

Finally we collect notification dates for conferences,
that is, the date when peer review was completed
and the results communicated to authors.
Notification dates were collected manually
from the conference web sites on 20 September 2017.
PLVDB is excluded from this analysis, as it has
a rolling deadline every month.
SIGMOD 2017 had an opportunity for revision during the review process; the date of the final accept/reject decision was used as the notification deadline.
In some cases, the date of notification was unavailable from a web search:
\begin{itemize}
\tightlist
\item For ICML 2017, the notification date was taken from the date of the author notification email that the first author received from his submission to the conference. 
\item For SODA 2017, the notification deadline was taken to be 1 October 2016 as an approximation based on a statement made at \url{https://www.siam.org/meetings/da17/proceedings.php}: "SIAM will provide final paper submission instructions to authors of accepted papers in early October 2016". 
\item For ICRA 2017, the notification date was unavailable online so the camera-ready date was used instead.
\end{itemize}

All code and data required to reproduce
this analysis is freely available at {\url{http://groups.inf.ed.ac.uk/cup/csarxiv/}},
with figures generated using Jupyter Notebooks
(\url{http://jupyter.org/}).

\subsection{Threats to Validity}

There are many types of e-print servers 
that this study does not attempt to address,
e.g., other centralized e-print servers than arXiv, institutional e-print repositories, and
dissemination via authors' web sites. 
Perhaps the most important centralized e-print 
repositories in computer science
that are not considered are 
bioRxiv (\url{biorxiv.org/}),
HAL (\url{https://hal.archives-ouvertes.fr/}),
the Electronic Colloquium on Computational Complexity (\url{http://www.eccc.uni-trier.de/}), 
and the Cryptology ePrint archive
(\url{http://eprint.iacr.org/}).
In this study, we focus on arXiv because of its
popularity, as 
we are unaware of another repository that is as
popular in computer science, and because ACM
CoRR is hosted via arXiv.org.
However, it is therefore important to note that these data necessarily provide
an underestimate of e-print usage in computer science.
Furthermore, our results are also an underestimate because
our record linkage procedure is intentionally conservative.

It is only possible to obtain metadata for
papers published in conferences rather than
all submissions, because for the majority of
conferences, the review process is confidential.
So it is not possible for us to measure
whether e-print usage among accepted papers
is different from rejected papers, e.g. because
of differences at an aggregate level in the population
of authors.
 
\begin{figure}
\centering
	\includegraphics[width=\textwidth]{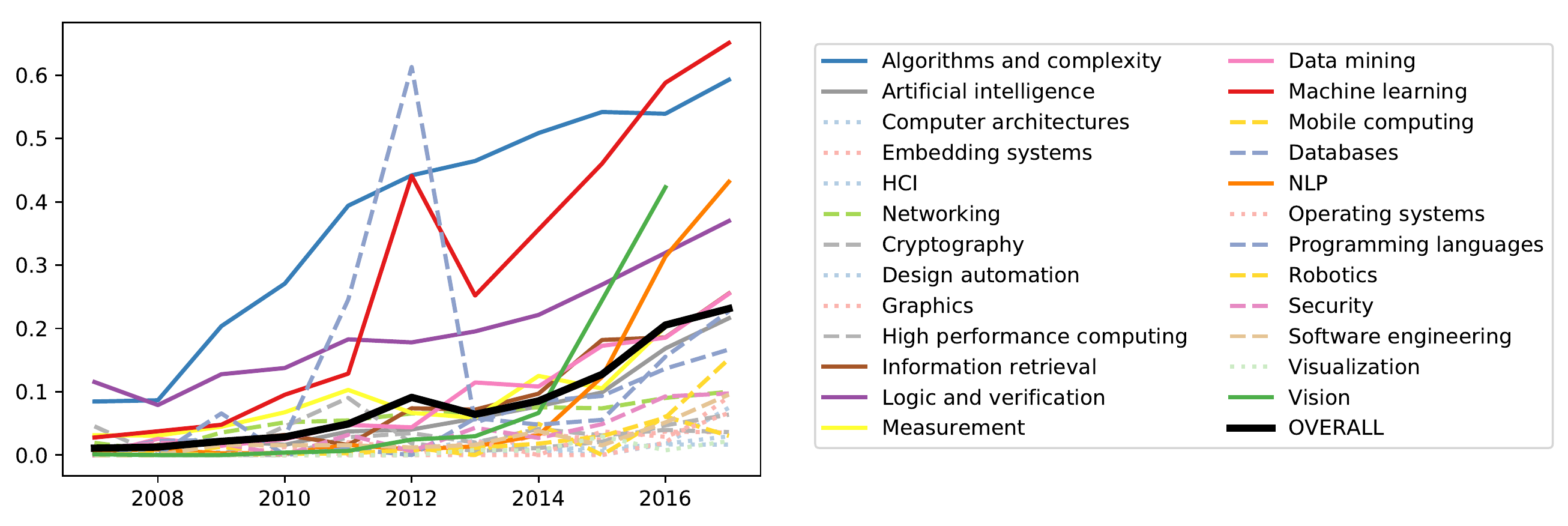}
	\caption{Proportion of papers in \nconf top
	conferences in computer science that have e-prints
	on arXiv.org,
	by area.}\label{fig:pct}
\end{figure}

At the time of writing, around a third of the research conferences considered in this study
have not yet uploaded publication data for
2017 to DBLP. 
For these conferences, we have included pre-2017
data in Figure~\ref{fig:pct}, but  not 
in the other figures. We will update the
e-print of this paper
as that information becomes available.

\section{Results}\label{results}

\begin{table}
\centering
\begin{tabular}{lrrr@{\%\hspace{1em}}rr@{\%}}
\toprule
       Venue &  Papers &  
              \multicolumn{2}{c}{e-prints} &
              \multicolumn{2}{c}{Preprints} \\
        & & 
        {\small Count} & \multicolumn{1}{c}{\small (\%)} &
        {\small Count} & \multicolumn{1}{c}{\small (\%)} \\

\midrule
             ICML &     427 &       278 &           65 &        204 &            48 \\
             STOC &      98 &        63 &           64 &         49 &            50 \\
             SODA &     182 &       103 &           57 &         81 &            45 \\
              ACL &     290 &       133 &           46 &         26 &             9 \\
            EMNLP &     303 &       123 &           41 &         45 &            15 \\
             LICS &      85 &        33 &           39 &         22 &            26 \\
              CAV &      61 &        21 &           34 &         10 &            16 \\
             POPL &      64 &        19 &           30 &         10 &            16 \\
              WWW &     158 &        44 &           28 &         24 &            15 \\
              KDD &     208 &        53 &           25 &         35 &            17 \\
             AAAI &     661 &       165 &           25 &         82 &            12 \\
            SIGIR &      77 &        16 &           21 &          0 &             0 \\
            IJCAI &     653 &       119 &           18 &         51 &             8 \\
             ICRA &     771 &       117 &           15 &         84 &            11 \\
           SIGMOD &     107 &        15 &           14 &          8 &             7 \\
              FSE &      90 &        10 &           11 &          0 &             0 \\
 SIGGRAPH (+Asia) &     163 &        15 &            9 &          0 &             0 \\
\bottomrule
\end{tabular}
	
\caption{Proportion of e-prints and preprints
of number of total accepted papers in 2017.
The conferences shown are those that had at least 10 e-prints
on arXiv.org in 2017.}\label{tbl:top_venues}
\end{table}

\begin{figure*}
	\centering
	\includegraphics[width=\textwidth]{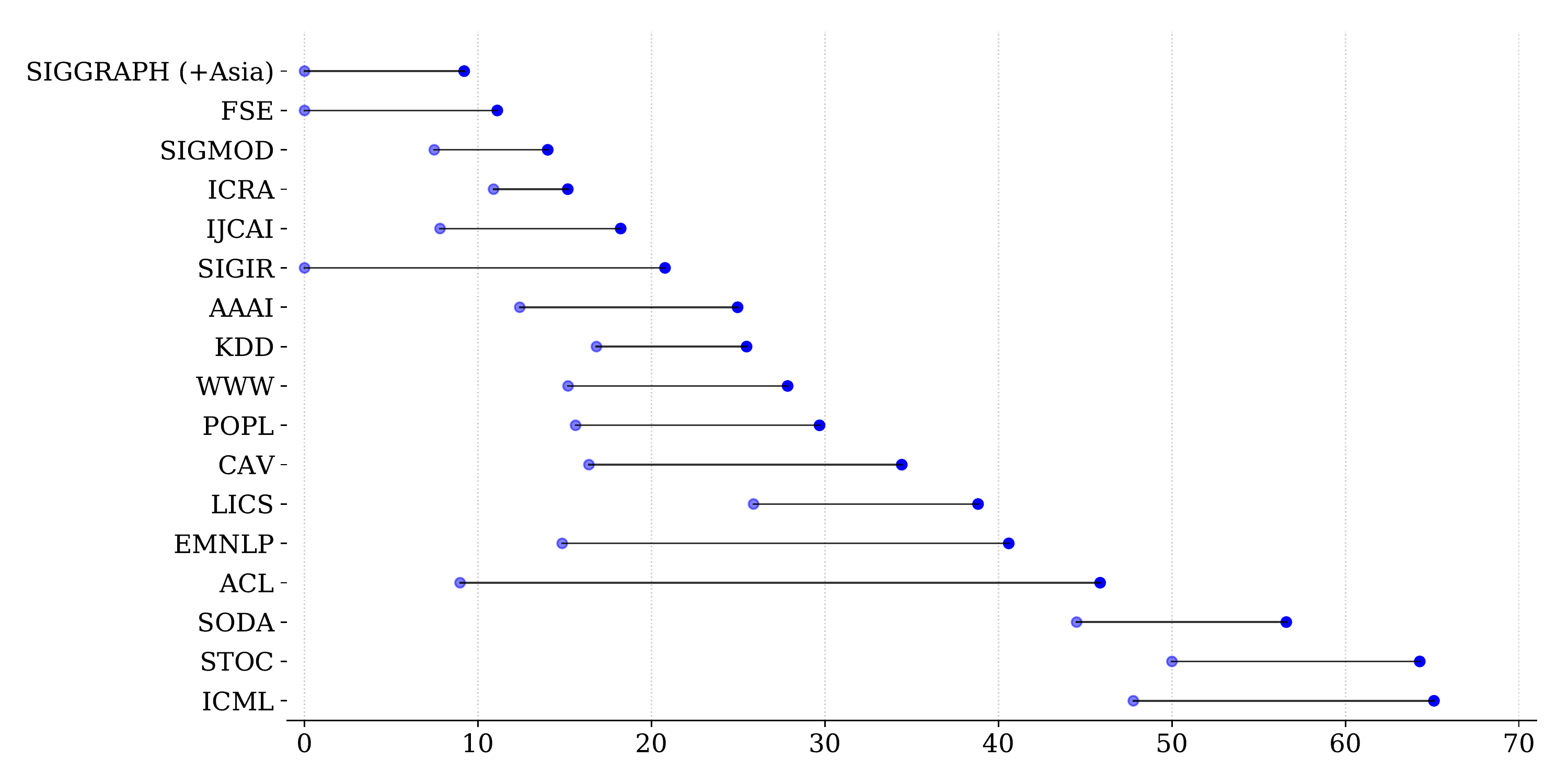}
	\caption{Visualization of e-print and preprint percentages in 2017. For each conference, the
	darker blue circles on the right indicate e-print percentage,
	and the lighter blue circles on the left indicate preprint percentage. The conferences shown are those that had at least 10 e-prints
on arXiv.org in 2017.}
\end{figure*}
\begin{figure*}[p]
	\centering
	\includegraphics[width=\textwidth]{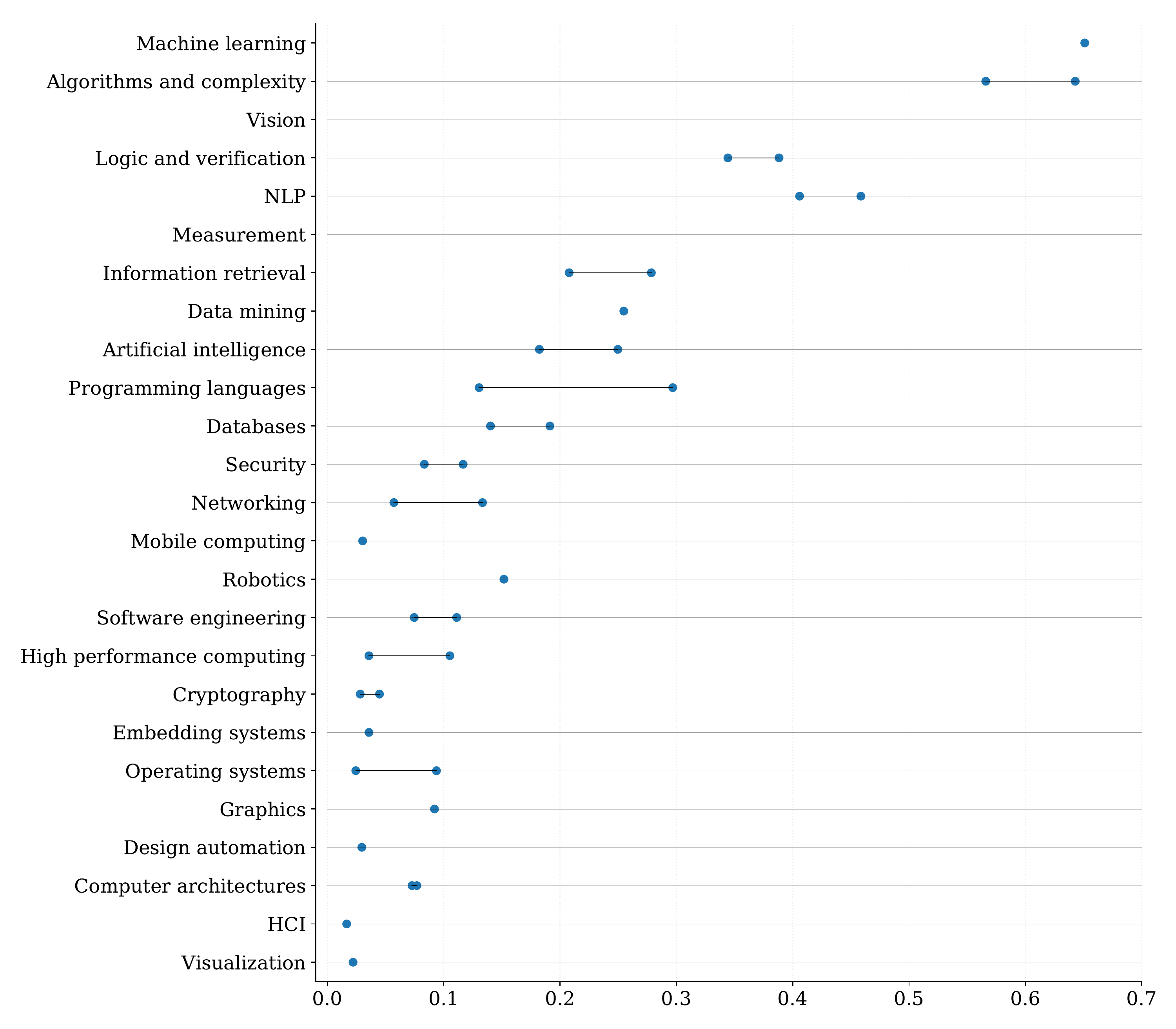}
	\caption{Proportion of e-prints
	for each conference in 2017, arranged by area.
	This shows the variation in e-print prevalence
	within each area. Some conferences considered
	in this study are missing because they have
	not yet uploaded publication data to DBLP.}\label{fig:dotplot}
\end{figure*}

We show the percentage of published papers per
year that have
arXiv e-prints in Figure~\ref{fig:pct}.
The ``Overall'' line is a microaverage across areas,
that is, the total number of e-prints across all conferences divided by the total number of publications.
We see that the total percentage of e-prints
has risen dramatically over the past decade, 
increasing from under 1\% in 2007 to 23\% in 2017.
It can also be seen that there is large variation
in e-print percentage across different areas. Machine learning and algorithms/complexity have the highest percentage of
e-prints, while NLP and computer vision have seen
a sharp increase since 2014. Logic and verification
is the final area whose e-print percentage
is much above the microaverage; although in 2007,
it had the highest e-print rate across computer
science, since
then usage has grown more slowly than other areas.
Most other areas are below the overall
average, but a large number of areas have
increased over the past two or three years.

In addition to examining variation across areas
of computer science, we also examine whether
there is variation in arXiv usage within the 
conferences in a research area. 
As Figure~\ref{fig:dotplot} shows, 
the variation in e-print percentage within an 
area seems to be much less than the variation
across areas.

The previous analysis has measured percentage of e-prints, 
that is, whether published papers are on arXiv at all,
but it does not measure whether those papers are \emph{preprints}, that is, whether
they were posted publicly before they were 
accepted into the conference. To measure this,
we collected the list of conference submission deadlines and notification dates for all \ntopconf
conferences which had at least 10 e-prints on arXiv
in 2017. These conferences are listed in Table~\ref{tbl:top_venues}. We also collected the \emph{notification
date} of each conference, that is, the date
at which authors were informed of the review outcome, 
as described in the previous section.
This allows us to measure for each conference,
what percentage of the published papers were 
preprints.

The results are shown in Table~\ref{tbl:top_venues}.
This table shows the number of papers published
at each venue, along with what percentage appear
as e-prints, and what
percentage appear as preprints, i.e., the date of 
the arXiv submission is before the conference
notification date.
Overall 56\% of e-prints on arXiv were also preprints, a rate which seems to approximately
hold across conferences, with a few exceptions.
The three conferences with the most e-prints (ICML,
STOC, and SODA) have a larger percentage of preprints, with around \emph{50\% of published papers being preprints}. In the other direction,
ACL and SIGIR have much lower rates of preprints
than one would expect given their percentages
of e-prints. This indicates that even
among communities in which arXiv e-prints
are widely accepted, there is still substantial
disagreement about preprint
norms.

\section{Conclusions}\label{policy}
\label{conclusions}

Overall, these data seem to indicate
that there is 
openness for change across computer science
in how research results are disseminated.
Prepublication and centralized e-print repositories
have reached a level of popularity in computer
science that it seems fair to say that they are here to stay,
but the fact that this trend has happened
relatively recently indicates that communities
may still be willing to experiment.
Researchers and practitioners need to think
  about how to respond in order to maximize the benefits and minimize
  the down-sides of this model.
We suggest different implications for different
communities that produce, read, and apply
research in computer science:

\begin{itemize}
\item
  \emph{Authors} should consider how
centralized e-print repositories fit into their dissemination
plans. Considering the growing popularity
of arXiv/CoRR, and the advantages of centralized
e-print repositories compared to web pages
of individual research groups,
we would argue that all research
papers in computer science should have public e-prints
on arXiv/CoRR,
unless a more specialized repository
is already popular in  the area.
\item
  \emph{Reviewers, journal editors, and conference
  chairs} for venues that use double-blind reviewing need to be aware of
  the increasing prevalence of preprints,
  and adjust reviewing guidelines and norms accordingly.
There have always been ways that reviewers
could become deblinded, learning the author identities
of a double-blind submission
by seeing previous technical reports, workshop
presentations, and invited lectures by the authors
on similar topics.
But centralized e-print repositories increase
the risk of reviewer deblinding, as repositories 
send alerts
to large numbers of researchers when new papers
are added.

Reviewing norms for double-blind venues
should be adjusted to reduce the negative
impact of deblinding.
This includes reviewers refraining where possible
from Web searches
that could deblind the authors, reviewers notifying conference
chairs when they have become deblinded,
and conference chairs being prepared to downweight
reviewer recommendations or reassign 
reviewers in such cases.

\item \emph{Research communities} need to have
a long-term discussion about whether or not to 
encourage preprints.
Although we argue strongly for centralized e-prints,
whether those should be preprints is a more difficult
question, because of the interaction with double-blind review.
On the one hand, double-blind review has been demonstrated to benefit
researchers from underrepresented communities,
while on the other hand, centralized e-print
repositories and preprints also benefit
such researchers by increasing the flow of information both to and from unfairly
marginalized communities.
We are hopeful that the changes to reviewing norms
described above may help to reconcile potential
conflicts between preprints and double-blind review
--- indeed, even ICLR, which has innovated open review among
large conferences, is moving to double blind in 2018.

\item
  \emph{Practitioners} in computer science
should be aware that centralized e-print repositories
are now becoming a more important vehicle
for rapid dissemination across computer science,
and should consider subscribing to email alerts
from the repositories that are appropriate to
their area.

\item
\emph{Builders of tools} to support new
reviewing, dissmenination, and publication models
for computer science should be encouraged to redouble
their efforts by the research community's manifest
willingness to consider new models of publication.
In particular, this may be a good moment
to reconsider the introduction of arXiv overlay journals for computer science,
which have been proposed earlier by several
authors \citep{halpern2000corr,lecun:manifesto}.
  \end{itemize}
  
To fully understand the tradeoffs involved
in centralized e-print repositories, prepublication,
and reviewing models,
more information is needed than is available 
from these data. In particular, we have little
information about the detailed workings of
double-blind reviewing processes
across computer science, e.g., in areas where there
is a high rate of pre-prints, how often does this
cause author identities to be revealed in double-blind processes? Some researchers have suggested
anonymous preprints as a way of reconciling preprint with double-blind review\footnote{\url{https://chairs-blog.acl2017.org/2017/03/02/arxiv-and-double-blind-reviewing-revisited/}}, but this
could cause issues with papers that are repeatedly submitted unsuccessfully\footnote{\url{http://www.theexclusive.org/2017/09/arxiv-double-blind.html}}; how often are rejected papers 
resubmitted, and where? 
On the more general subject of learning more about the double-blind process,
a fascinating recent study is 
by \cite{legoues17blind}, who show that most
reviewers are unable to guess author identities
in double blind conference at three conferences
in programming languages and software engineering; it would be interesting
to see if this type of result can be replicated
at conferences with higher rates of preprints.
Because the reviewing process
of most conferences and journals is confidential,
it can be difficult to gather data,
but we hope that such studies have potential
to add light to the existing
heat of discussions of reviewing models
in the research community.

\section*{Acknowledgements}

We thank the teams behind arXiv.org and DBLP for their hard work both
in building useful services and in making their data publicly available, as well as Emery Berger
for the CSRankings service.
We thank Emery Berger, Kyunghyun Cho, and Paul Ginsparg for useful
discussions.

\bibliographystyle{plainnat}
\bibliography{cs-arxiv-popularity}

\begin{thebibliography}{13}
\providecommand{\natexlab}[1]{#1}
\providecommand{\url}[1]{\texttt{#1}}
\expandafter\ifx\csname urlstyle\endcsname\relax
  \providecommand{\doi}[1]{doi: #1}\else
  \providecommand{\doi}{doi: \begingroup \urlstyle{rm}\Url}\fi

\bibitem[ope()]{openreview:web}
Openreview.
\newblock \url{http://openreview.net/}.
\newblock Accessed September 2017.

\bibitem[Berger(2017)]{csrankings:web}
Emery Berger.
\newblock {CSRankings}.
\newblock \url{http://csrankings.org/}, 2017.
\newblock Accessed September 2017.

\bibitem[Giles et~al.(1998)Giles, Bollacker, and Lawrence]{giles98citeseer}
C.L. Giles, K.~Bollacker, and S.~Lawrence.
\newblock Citeseer: An automatic citation indexing system.
\newblock In \emph{DL'98 Digital Libraries, 3rd ACM Conference on Digital
  Libraries}, pages 89--98, 1998.

\bibitem[Ginsparg(2011)]{ginsparg11twenty}
Paul Ginsparg.
\newblock It was twenty years ago today ..
\newblock \emph{CoRR}, abs/1108.2700, 2011.
\newblock URL \url{http://arxiv.org/abs/1108.2700}.

\bibitem[Halpern(2000)]{halpern2000corr}
Joseph~Y Halpern.
\newblock {CoRR}: a computing research repository.
\newblock \emph{ACM Journal of Computer Documentation (JCD)}, 24\penalty0
  (2):\penalty0 41--48, 2000.

\bibitem[Kling(2005)]{kling05}
R~Kling.
\newblock The internet and unfrefereed scholarly publishing.
\newblock \emph{Annual Review of Information Science and Technology},
  38\penalty0 (1):\penalty0 591--631, 2005.

\bibitem[Larivi\`ere et~al.(2014)Larivi\`ere, Sugimoto, Macaluso,
  Milojevi{\'c}, Cronin, and Thelwall]{lariviere14}
V.~Larivi\`ere, C.~R. Sugimoto, B.~Macaluso, S.~Milojevi{\'c}, B.~Cronin, and
  M.~Thelwall.
\newblock arxiv e-prints and the journal of record: An analysis of roles and
  relationships.
\newblock \emph{Journal of the Association for Information Science and
  Technology}, 65:\penalty0 1157--1169, 2014.

\bibitem[Lawrence et~al.(1999)Lawrence, Giles, and
  Bollacker]{lawrence1999autonomous}
Steve Lawrence, C~Lee Giles, and Kurt~D Bollacker.
\newblock Autonomous citation matching.
\newblock In \emph{Conference on Autonomous Agents}, pages 392--393. ACM, 1999.

\bibitem[{Le Goues} et~al.(2017){Le Goues}, Brun, {Apel}, {Berger}, {Khurshid},
  and {Smaragdakis}]{legoues17blind}
Claire {Le Goues}, Yuriy Brun, Sven {Apel}, Emery {Berger}, Sarfraz {Khurshid},
  and Yannis {Smaragdakis}.
\newblock {Effectiveness of Anonymization in Double-Blind Review}.
\newblock \emph{ArXiv e-prints}, September 2017.
\newblock URL \url{https://arxiv.org/abs/1709.01609}.

\bibitem[{LeCun}(2009)]{lecun:manifesto}
Yann {LeCun}.
\newblock A new publishing model in computer science, 2009.
\newblock URL \url{http://yann.lecun.com/ex/pamphlets/publishing-models.html}.
\newblock Accessed September 2017.

\bibitem[McCallum et~al.(2000{\natexlab{a}})McCallum, Nigam, Rennie, and
  Seymore]{mccallum:cora}
Andrew McCallum, Kamal Nigam, Jason Rennie, and Kristie Seymore.
\newblock Automating the construction of internet portals with machine
  learning.
\newblock \emph{Information Retrieval Journal}, 3:\penalty0 127--163,
  2000{\natexlab{a}}.

\bibitem[McCallum et~al.(2000{\natexlab{b}})McCallum, Nigam, and
  Ungar]{mccallum2000efficient}
Andrew McCallum, Kamal Nigam, and Lyle~H Ungar.
\newblock Efficient clustering of high-dimensional data sets with application
  to reference matching.
\newblock In \emph{International Conference on Knowledge Discovery and Data
  Mining (KDD)}, pages 169--178. ACM, 2000{\natexlab{b}}.

\bibitem[Soergel et~al.(2013)Soergel, Saunders, and McCallum]{openreview}
David Soergel, Adam Saunders, and Andrew McCallum.
\newblock Open scholarship and peer review: a time for experimentation.
\newblock In \emph{ICML Workshop on Peer Reviewing and Publishing Models
  (PEER)}, 2013.

\end{thebibliography}

\end{document}